\documentclass[jkps,preprint,fleqn,showpacs,showkeys]{revtex4}
\usepackage[pdftex]{graphicx}
\usepackage{amssymb}
\usepackage{amsmath}
\usepackage{bm}

\newcommand\xHI{x_{\rm HI}}
\newcommand\avgxHI{\bar{x}_{\rm HI}}
\newcommand\avgxHIsub[1]{\bar{x}_{\rm HI,#1}}
\newcommand\dTb{\delta T_{\rm b}}
\newcommand\dTbrms{\delta T_{\rm b,rms}}
\newcommand\Mpc{\mathinner{\mathrm{Mpc}}}
\newcommand\mK{\mathinner{\mathrm{mK}}}
\renewcommand\d{\mathrm{d}}
\renewcommand\vec[1]{\bm{\mathbf{#1}}}
\newcommand\Msun{\mathinner{\mathrm{M}_\odot}}
\newcommand\MHz{\mathinner{\mathrm{MHz}}}
\newcommand\removed[1]{}

\def\apjl{Astrophys. J. Lett.}
\def\mnras{Mon. Not. R. Astron. Soc.}
\def\nat{Nature}
\def\aap{Astron. Astrophys.}
\def\apj{Astrophys. J.}
\def\aj{Astron. J.}

\def\prd{Phys. Rev. D}

\def\physrep{Physics Reports}

\def\pasa{Publ. Astron. Soc. Aust.}
\def\pasp{Publ. Astron. Soc. Pac.}
\def\araa{Annu. Rev. Astron. Astrophys.}

\usepackage{multirow}

\begin{document}
\setcounter{page}{0}
\title[Deep Learning for EoR]{Deep-Learning Study of the 21cm Differential Brightness Temperature During the Epoch of Reionization}
\author{Yungi \surname{Kwon}}
\affiliation{Department of Physics, University of Seoul,
163 Seoulsiripdaero, Dongdaemun-gu, Seoul 02504, Republic of Korea}
\author{Sungwook E. \surname{Hong}}
\email{swhong83@uos.ac.kr}
\affiliation{Natural Science Research Institute, University of Seoul,
163 Seoulsiripdaero, Dongdaemun-gu, Seoul 02504, Korea}
\author{Inkyu \surname{Park}}
\affiliation{Department of Physics, University of Seoul,
163 Seoulsiripdaero, Dongdaemun-gu, Seoul 02504, Republic of Korea}
\affiliation{Natural Science Research Institute, University of Seoul,
163 Seoulsiripdaero, Dongdaemun-gu, Seoul 02504, Korea}

\date[]{Received \today}

\begin{abstract}
We propose a deep learning analyzing technique with convolutional neural network (CNN) to predict the evolutionary track of the Epoch of Reionization (EoR) from the 21-cm differential brightness temperature tomography images. 
We use 21cmFAST, a fast semi-numerical cosmological 21-cm signal simulator, to produce mock 21-cm maps between $z=6 \sim 13$.
We then apply two observational effects into those 21-cm maps, such as instrumental noise and limit of (spatial and depth) resolution somewhat suitable for realistic choices of the Square Kilometre Array (SKA).
We design our deep learning model with CNN to predict the sliced-averaged neutral hydrogen fraction from the given 21-cm map.
The estimated neutral fraction from our CNN model has a great agreement with its true value even after coarsely smoothing with broad beamsize and frequency bandwidth, and also heavily covered by noise with narrow.
Our results have shown that deep learning analyzing method has a large potential to efficiently reconstruct the EoR history from the 21-cm tomography surveys in future.
\end{abstract}



\maketitle

\section{Introduction} \label{sec:intro}

About 380,000 years after the Big Bang, the Universe had cooled down enough that the floating free protons and electrons could combine to form the neutral hydrogen atoms (HI). As there were no luminous objects such as stars and galaxies, this period is referred to as the cosmic dark ages. During the cosmic dark ages, most of the hydrogen atoms exist as neutral. 
Then, by the gravitational collapse of the overdense regions in the intergalactic medium (IGM), they started to form more and more pronounced structures of the Universe. Eventually, the first luminous objects started to form, which leads the epoch of reionization (EoR). During this epoch, the first radiating objects in the Universe heat and re-ionize the surrounding local neutral IGM by energetic radiation. 

As the EoR involves the key science to understand the formation of first objects and large-scale structures, numerous studies have performed to understand the epoch. For example, several observations have constrained the duration of the EoR from redshifts $z \simeq 12$ to $6$ \cite{c1,c3,c2,c4,c5}. However, due to the lack of bright objects, the current observational evidences of the EoR are rather limited. For instance, we poorly understand for what is the main source governing the EoR and what properties lead the ionizing process, and how did it intimately affect on subsequent structure formation. 

Several observational methods have been suggested to understand the EoR.
For example, \cite{mortonson2008, ahn2012, heinrich2017} studied how the cosmic microwave background (CMB) anisotropy observations can constrain the evolutionary history of the mean neutral hydrogen fraction ($\avgxHI (z)$).
Among various observational methods for studying the EoR,
one of the most promising ways is to use the 21-cm wavelength radiation emitted by the hyperfine transition of neutral hydrogen atom, as the hydrogen gas was predominant component of the Universe. Several radio telescope experiments are being planned and accumulated for observing the redshifted 21-cm signal from the EoR: the Murchison Widefield Array (MWA)\cite{c6}, the Giant Metrewave Radio Telescope (GMRT)\cite{c7}, the Low Frequency Array (LOFAR)\cite{c8}, the Precision Array Probing the Epoch of Reionization (PAPER) \cite{c9}, the Hydrogen Epoch of Reionization Array (HERA)\cite{c10}, and the Square Kilometre Array (SKA)\cite{c11, c12}.  
Nevertheless, there remains a challenge to extract the proper redshifted 21-cm signal during the EoR from radio observations, due to the foregrounds from the extra-galactic radio sources and diffuse galactic foreground, the ionospheric distortions, and the instrumental noise, etc. Specifically, the Galactic synchrotron radiation, which has similar wavelength to those of the redshifted 21-cm signal during the EoR, is expected to be much brighter than the cosmological 21-cm signal\cite{c14, c13, c15}. 

While numerous methods have suggested on subtracting foregrounds and analyzing the 21-cm signals\cite{c16,c18,c19,c17}, brand-new analyzing techniques have increasingly approved recently. 
One of them is the deep learning, a subset of machine learning that works with artificial neural networks (ANN), that is designed to imitate how humans think and learn by using the so-called neurons. 
The neural network accepts a series of data as input and grasps the certain patterns within the data by fitting the weights on the connections between neurons of the network. 
Deep learning technique has shown to have enormous potential on astrophysics\cite{c20,c21,c22,c23,list2020,hortua2020,laplante2019}. For example, \cite{c24} used the deep learning technique to estimate the ionizing efficiency, the minimum virial temperature of halos, and the mean free path of ionizing photons by using the 21-cm power spectra at different redshifts as inputs. Also, \cite{c25} used the entire 21-cm lightcone map as the input of the convolutional neural network (CNN) to discriminate the re-ionizing process from the active galactic nuclei (AGNs) and star-forming galaxies. To date, however, a series of studies has proceeded to deal with somewhat ideal simulated data as being poorly considered the actual observational effects.

In this paper, we introduce a novel deep learning-based method by re-constructing $\avgxHI(z)$ during the EoR from mock 21-cm differential brightness temperature maps by considering various observational effects. We simulate 21-cm maps by using a semi-numerical simulation code 21cmFAST\cite{c26} at different redshifts. We convolve them with certain beamsize and frequency bandwidth under somewhat realistic observational conditions, and add the corresponding white Gaussian noise to generate the noisy maps. We use those noisy 21-cm maps as inputs of our CNN model and train the model to predict the corresponding value of $\xHI$.  

This paper is organized as follows. In Section~\ref{sec:2}, we describe how to simulate the mock 21-cm differential brightness temperature maps. In Section~\ref{sec:3}, we introduce our CNN architecture and training method. In Section~\ref{sec:4}, we show the performance of our method with various observational conditions. Finally, in Section~\ref{sec:5}, we summarize our results. 

Throughout this paper, we adopt the background cosmological parameters best fit to the standard $\Lambda$CDM cosmology from Planck 2018\cite{c48}:
matter density parameter $\Omega_{\rm m} = 0.31$,
baryon density parameter $\Omega_{\rm b} = 0.048$,
cosmological constant density parameter $\Omega_{\Lambda} = 0.69$,
spectral index $n_s = 0.97$,
rms density fluctuation $\sigma_8 = 0.81$,
Hubble parameter $h = 0.68$,
and the primordial helium abundance $Y_p = 0.245$.

\section{Simulation of Noisy Mock 21-cm Maps}\label{sec:2}

\subsection{Simulating Undistorted Redshifted 21-cm Signals}

We use the 21cmFAST\cite{c26} to generate the 21-cm signal and corresponding neutral fraction with the redshifts $z = 6 \sim 13$, where most of the current studies strongly have constrained. The 21cmFAST is the publicly-available simulation of the cosmological reionization which is self-consistent and semi-numerical. Specifically, this is optimized to generate the 21-cm signal during the epoch. By the combinations of the excursion-set formalism\cite{c42, c35} which mentioned earlier for identifying ionized hydrogen regions and the first-order perturbation theory\cite{c39}, it generates the full three-dimensional realizations of the density, ionization field, velocity, and spin temperature and finally the 21-cm differential brightness temperature with a given redshift.  

We first generate the density and velocity fields with $50$, $100$, $200$, and $300\Mpc$ comoving boxsizes and $800^3$, $1000^3$, and $1250^3$ grids between $z = 6 \sim 13$.
We use different random seeds to generate the initial conditions for each combination of comoving boxsize and grid number, and, as a result, we get 12 independent simulation sets.
For each simulation set, we obtain $18\sim 35$ snapshots between $z = 6\sim 13$, where the number of snapshots depends on the number of grids.
Note that the spatial resolutions used in this paper ($\Delta x = 0.04 \sim 0.375 \Mpc$) are smaller than the length scale where density perturbation from the 21cmFAST agrees well with the $N$-body simulations\cite{c26}.
However, since we do not focus on small-scale details and the output 21-cm maps would be smoothed in larger scales, our choice of spatial resolution does not severely affect our result.

Then we calculate the ``undistorted'' 21-cm differential brightness temperature, $\dTb^0 (\vec{x},z)$, by downgrading the resolution to $200^3$ grids: 
\begin{equation}
\dTb^0 \approx (27 \mK) \xHI (1+\delta) \left( \frac{H}{\d v_r/\d r + H} \right) \left(1 - \frac{T_{\rm CMB}}{T_{\rm S}} \right) \left(\frac{1+z}{10} \frac{0.15}{\Omega_{\rm m} h^2} \right)^{1/2} \left( \frac{\Omega_{\rm b} h^2}{0.023} \right) \, ,
\label{eq:dTb}
\end{equation}
where $\delta$, $T_{\rm CMB}$, and $T_{\rm S}$ are the gas overdensity, the CMB temperature, and the gas spin temperature, respectively\cite{c13}.
In 21cmFAST, $T_{\rm S}(z)$ is calculated from the evolution of the kinetic gas temperature and the Ly-$\alpha$ background.
Also, $\xHI(\vec{x},z)$ is calculated from the number of IGM ionizing photons per baryon \cite{park2019}
\begin{equation}
n_{\rm ion} = \bar{\rho_{\rm b}}^{-1} \int_{0} ^{\infty} \d M_{\rm h} \frac{\d n(M_{\rm h},z)}{\d M_{\rm h}} f_{\rm duty} M_{\star} f_{\rm esc} N_{\gamma/{\rm b}} \, ,
\end{equation}
where $\bar{\rho_{\rm b}}$, $M_{\rm h}$, $f_{\rm duty}$, $M_{\star}$, $f_{\rm esc}$, $N_{\gamma/{\rm b}}$ are the mean baryon density, halo mass, factor related to the suppression of star formation at massive halos, stellar mass, escape fraction, and the number of ionizing photons per stellar baryon, respectively.
$f_{\rm duty}$, $M_{\star}$, and $f_{\rm esc}$ can be expressed as a function of $M_{\rm h}$ as follows:
\begin{align}
f_{\rm duty}(M_{\rm h}) &= \exp \left( -\frac{M_{\rm h}}{M_{\rm turn}} \right) \\
M_{\star} (M_{\rm h}) &= f_{\star} \frac{\Omega_{\rm b}}{\Omega_{\rm m}} M_{\rm h} = f_{\star,10} \left( \frac{M_{\rm h}}{10^{10} \Msun} \right)^{\alpha_{\star}} \frac{\Omega_{\rm b}}{\Omega_{\rm m}} M_{\rm h} \\
f_{\rm esc} (M_{\rm h}) &= f_{\rm esc,10} \left( \frac{M_{\rm h}}{10^{10} \Msun} \right)^{\alpha_{\rm esc}} \, ,
\end{align}
where $M_{\rm turn}$ is the halo mass threshold for efficient star formation, and $f_\star$ is the fraction of galactic gas in stars.
In this paper, we adopt a fiducial setup in \cite{park2019} for reionization parameters: $(N_{\gamma/{\rm b}}, f_{\star,10}, \alpha_{\star}, f_{\rm esc,10}, \alpha_{\rm esc},M_{\rm turn}) = (5000, 0.05, 0.5, 0.1, -0.5, 5\times 10^8 \Msun)$.
Note that we do not test different values of reionization parameters in this paper, mainly because we focus on the evolution of the mean neutral hydrogen fraction $\avgxHI(z)$ rather than constraining the reionization model.

\begin{figure}
\includegraphics[width=0.6\textwidth]{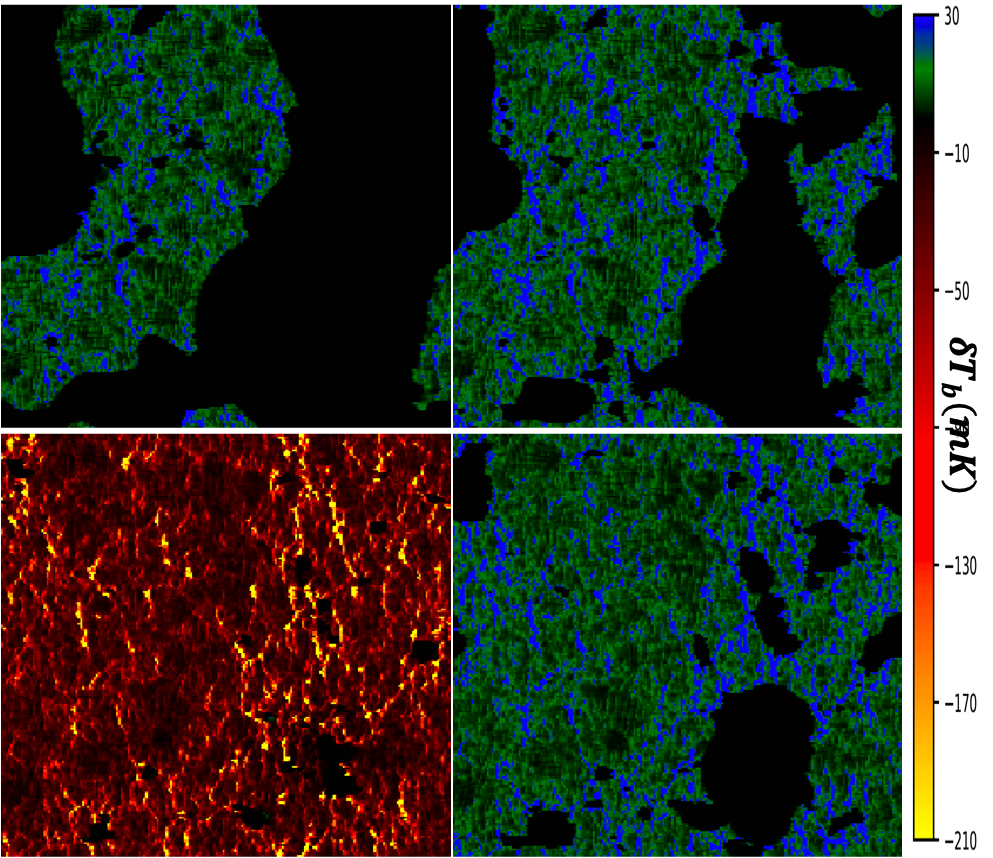}
\caption{An example of undistorted mock 21-cm maps with $50\,{\rm cMpc}$ length and $0.25\,{\rm cMpc}$ thickness.
From bottom-left to counter-clockwise direction:
$(z,\avgxHI) = (10.66, 0.94)$, $(8.58, 0.76)$, $(7.52, 0.54)$, and $(6.57, 0.25)$.}
\label{fig:2.2}
\end{figure}

Figure~\ref{fig:2.2} shows an example of undistorted 21-cm maps in a $50\,{\rm cMpc}$-box.
As expected in Eq.~(\ref{eq:dTb}) (if $T_{\rm S} \gg T_{\rm CMB}$, which is believed to be satisfied during the main reionization epoch), the undistorted 21-cm maps are proportional to the matter density and follow Gaussian distribution at high $(z, \avgxHI)$.
On the other hand, at low $(z, \avgxHI)$, the overdense region becomes fully ionized and shows no 21-cm signal, and therefore, the distribution is highly non-Gaussian.
Note that the estimation of $\avgxHI$ from the undistorted 21-cm maps could be rather straightforward, as it is simply the fraction of area where $\xHI(\vec{x})$ is close to zero.
However, the estimation of $\avgxHI$ may not be that straightforward if one adds the observational effects to the 21-cm signal, which we will see in the next section.

\subsection{Processing with the Observational Effects}
In practice, we expect that the observed signal would be very different with Figure~\ref{fig:2.2}, because the actual observational signal will be obtained by integrating the angle, frequency, and observation time. At this point, the size of the angle and frequency are the limit of spatial and depth resolution of the telescope. As the size of the angle or frequency of the telescope becomes larger, the observed signal becomes more smoothed.
Then the small-scale information is lost and the estimation of $\avgxHI$ similar to that we mentioned at the previous section may fail.
On the other hand, the smaller angle or frequency could contain the small-scale information. 
However, if the integration time is fixed, the level of noise becomes higher and the estimation of $\avgxHI$ could be also difficult.
Therefore, it is essential to include such observational effects, i.e., smoothing and noise, for the performance test of analysis methods such as our CNN model.

We start from smoothing the undistorted 21-cm maps with given angle and frequency setup of radio telescope.
Here, we use three choices of the beamsizes and frequency bandwidths---$\Delta \theta = 1'$, $2'$, and $3'$, and $\Delta \nu = 0.2\MHz$, $1\MHz$, and $2\MHz$.
The corresponding comoving length scales for given $\Delta \theta$ and $\Delta \nu$ are
\begin{align}
\Delta L_{\perp}(\Delta \theta, z) &= D_{\rm c}(z) \times \Delta \theta \\
\Delta L_{\parallel}(\Delta \nu, z) &= \frac{c (1+z)^2}{H_0 \sqrt{\Omega_{\rm m} (1+z)^3 + \Omega_\Lambda}} \frac{\Delta \nu}{\nu_0} \, ,
\end{align}
where $D_c(z)$ is the comoving distance, and $\nu_0 = 1420\MHz$ is the rest-frame frequency of the 21-cm line.
Note that the subscripts `$\perp$' and `$\parallel$' emphasize that the beamsize and frequency bandwidth affect the smoothing of $\dTb$ in directions perpendicular and parallel to the line-of-sight (LOS), respectively.

The actual shape of smoothing kernels in directions parallel and perpendicular to the LOS ($W(\nu; \Delta \nu)$ and $W(\theta; \Delta \theta)$, respectively) strongly depends on the configuration of radio telescope, even with the identical choice of $(\Delta \theta, \Delta \nu)$.
Especially, the shape of $W(\theta; \Delta \theta)$ (beam shape) could greatly vary depending on the antennae design.
It is known that some ``dirty'' beam shape such as the compensated Gaussian kernel might be possible\cite{mellema2006}, and in that case the understanding of the $\avgxHI(z)$ could be extremely difficult\cite{c49}.
Nevertheless, for simplicity and because the configuration of future surveys is not fixed, we assume the Gaussian kernel for both beam and frequency smoothing kernels in this paper.
We use $\Delta \theta$ and $\Delta \nu$, or identically, $\Delta L_{\perp}$ and $\Delta L_{\parallel}$, as the full-width half-maximum of the beam and frequency smoothing kernel.

Also, since interferometric radio observations might be insensitive to the large-scale fluctuation of $\dTb$, we subtract the mean of the smoothed 21-cm signal from the map so that the average of the subtracted 21-cm map becomes zero.
In summary, we obtain the ``smoothed'' 21-cm map (or, $\dTb^{\rm s}(\vec{\theta},\nu)$, where $\nu = \nu_0  / (1+z)$) as follows:
\begin{align}
\dTb^{\rm s}(\vec{\theta},\nu) &= \dTb'(\vec{\theta},\nu) - \langle \dTb^{\rm s'}(\vec{\theta'},\nu) \rangle_{\vec{\theta'}} \\
\dTb^{\rm s'}(\vec{\theta},\nu) &= \int \d \nu' \int \d \vec{\theta'} \, \dTb^0 (\vec{\theta'},\nu') W_{\rm G}(\left|\vec{\theta}-\vec{\theta'} \right| ; \Delta \theta) W_{\rm G}(\nu - \nu'; \Delta \nu) \, .
\end{align}
Here, the subscript `G' emphasizes the Gaussian kernel.

\begin{figure}
\includegraphics[width=0.8\textwidth]{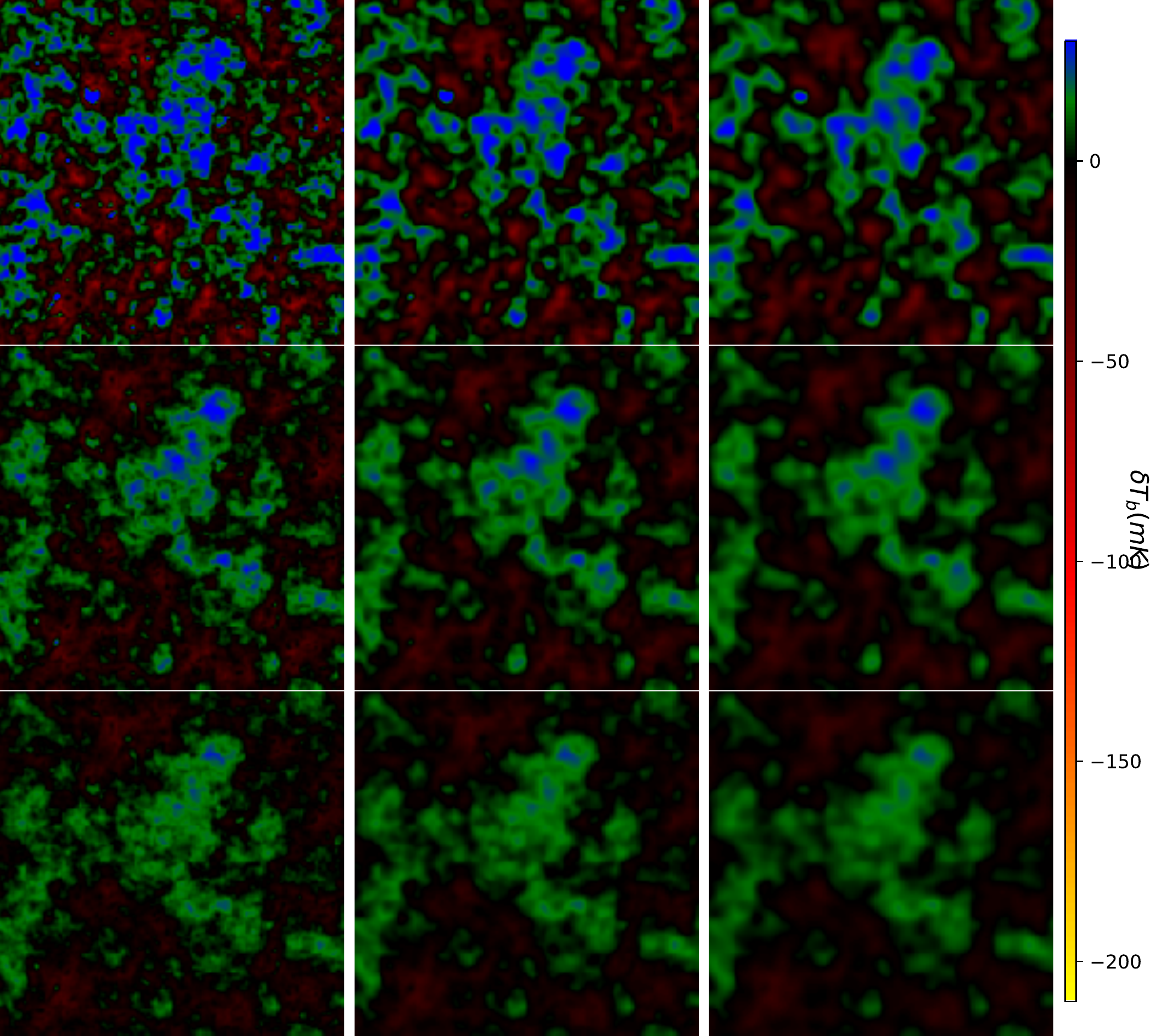}
\caption{An example of the ``smoothed'' mock 21-cm map in a $300\,{\rm cMpc}$-box at $(z,\avgxHI) = (12.73, 0.97)$. 
From left to right : beamsize $\Delta \theta = 1'$, $2'$, and $3'$. 
From top to bottom : frequency bandwidth $\Delta \nu = 0.2 \MHz$, $1 \MHz$, and  $2 \MHz$. 
We assume the Gaussian kernel for both beam and frequency smoothing.}
\label{fig:2.3}
\end{figure}

Figure~\ref{fig:2.3} shows an example of the smoothed mock 21-cm map in a $300\,{\rm cMpc}$-box at $(z,\avgxHI) = (12.73,0.97)$.
Compared to the undistorted 21-cm maps in Figure~\ref{fig:2.2}, there exists two key features in smoothed maps that make the estimation of $\avgxHI(z)$ more difficult.
First of all, due to the mean subtraction after smoothing, we cannot use the mean value ($\langle \dTb^{\rm s}(\vec{\theta},\nu) \rangle_{\vec{\theta}}$) as an indicator of $\avgxHI(z)$, especially for high-$z$ cases (see the bottom-left panel of Figure~\ref{fig:2.2} as a comparison).
Additionally, smoothing in both parallel and perpendicular directions to the LOS wipes out the small-scale features.
As a result, without a prior knowledge, the heavily smoothed 21-cm map at the lower-right panel of Figure~\ref{fig:2.3} might be confused to be at the middle of EoR (e.g., $\avgxHI(z) \sim 0.5$).
Note that, however, the spatial distribution of the smoothed 21-cm maps is Gaussian and highly non-Gaussian at $\avgxHI \simeq 1$ and $1-\avgxHI \gg 0$, respectively.
Therefore, it may be possible to distinguish between the smoothed 21-cm maps before and the middle of EoR by using $n$-point correlation functions or Minkowski functionals\cite{c49,wang2015}, while such methods usually require precise measurement of $\dTb$.

After smoothing, we add the noise to the mock 21-cm maps, whose level depends on the integration time and the configuration of radio telescope.
Here, we adopt the sensitivity limit at \cite{c49}, which uses the sensitivity calculation from \cite{iliev2003} by assuming the SKA with core size of $\sim 1\,{\rm km}$.
While the spatial and intensity distribution of noise and how it convolves with the signal also strongly depends on the configuration of radio telescope\cite{asorey2020}, here we adopt a somewhat simple model as follows.
We calculate the root-mean-square (RMS) of noise level ($\sigma(\Delta T; \Delta \theta, \Delta \nu)$) by using 10-, 100-, 1000-, and 10000-hour integration times and produce the Gaussian white noise with given RMS level and random seeds ($\epsilon$).
Then we obtain the ``noisy'' 21-cm map (or, $\dTb^{\rm n}(\vec{\theta}, \nu)$) by adding the noise to the smoothed 21-cm map,
\begin{equation}
\dTb^{\rm n}(\vec{\theta},\nu) = \dTb^{\rm s}(\vec{\theta},\nu) + \mathcal{W}_{\rm G}(\vec{\theta}; \sigma(\Delta T;\Delta \theta,\Delta \nu),\epsilon) \, .
\end{equation}
Note that the mean value of the noisy 21-cm map remains zero.

\begin{figure}
\includegraphics[width=\textwidth]{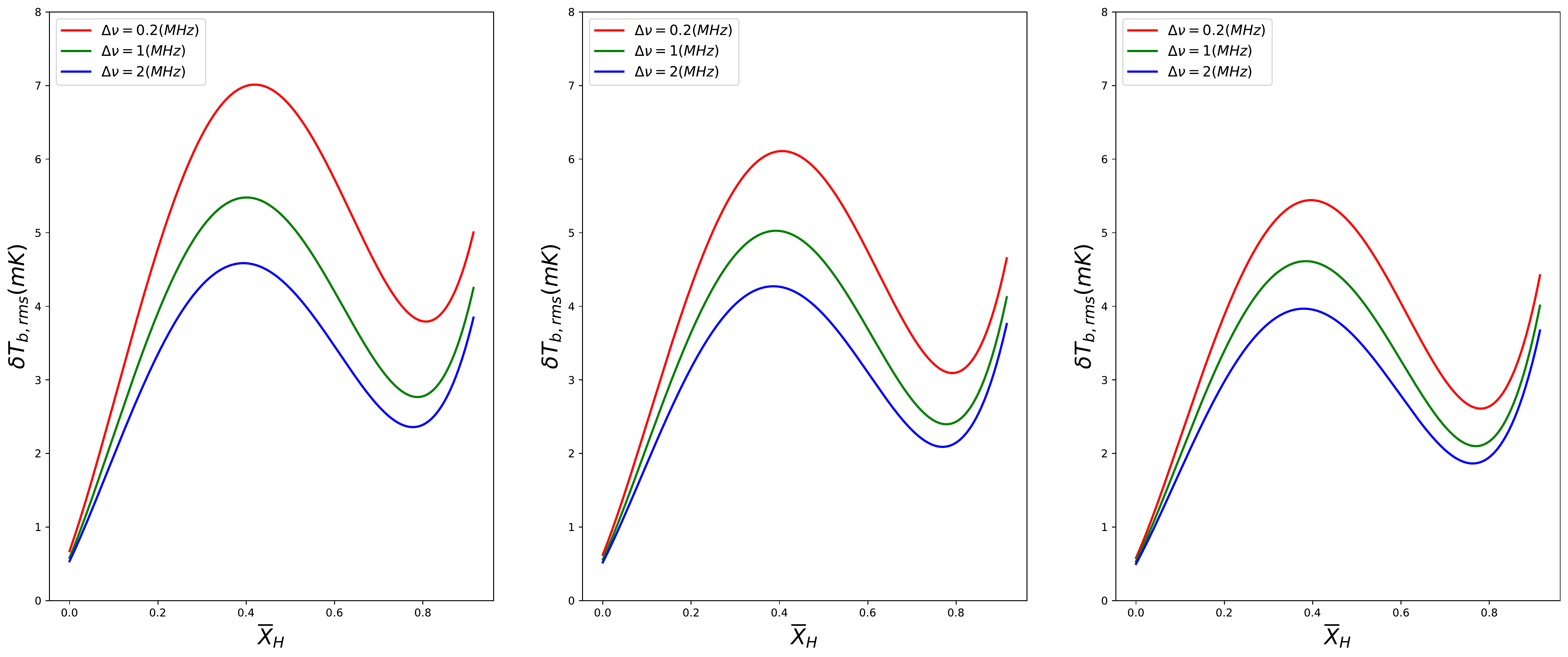} \\
\includegraphics[width=\textwidth]{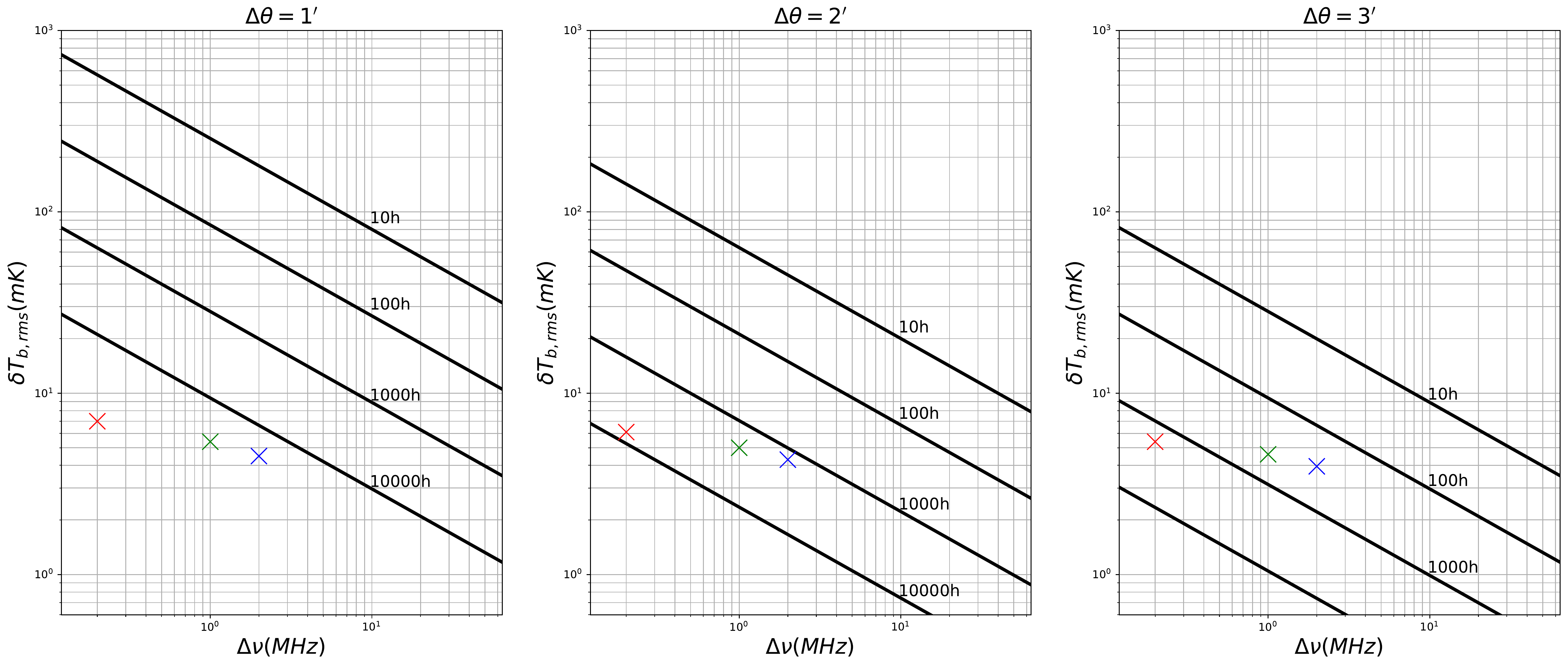}
\caption{Top panel: RMS of smoothed mock 21-cm maps as a function of mean neutral hydrogen fraction. 
Bottom panel: sensitivity limits with 10-, 100-, 1000-, and 10000-hour integration time of SKA (solid lines; from top to botom; \cite{c49}), compared to the maximum RMS of the smoothed mock 21-cm maps over time (peak value at the top panel; crosses).
From left to right: $\Delta \theta = 1'$, $2'$, and $3'$.
Colors: $\Delta \nu = 0.2\MHz$(red), $1\MHz$(green), and $2\MHz$(blue).}
\label{fig:2.5}
\end{figure}

Figure~\ref{fig:2.5} shows the comparison between our sensitivity limit estimation and the RMS value of the smoothed mock 21-cm maps ($\dTbrms$) with various configurations of beamsize, frequency bandwidth, and integration time.
Here, $\dTbrms / \sigma$ can be regarded as a proxy of the signal-to-noise ratio (SNR).
As commented earlier, to fully utilize the advantages of some $n$-points or topological analysis methods, one needs to achieve ${\rm SNR} \gg 1$. 
By assuming marginal choices of beamsize and frequency bandwidth, e.g., $\Delta \theta = 1' \sim 2'$ and $\Delta \nu = 1 \sim 2 \MHz$, such SNR is possible only with $\gtrsim 10000$-hour SKA integration time\cite{c49}.

\begin{figure}
\includegraphics[width=0.8\textwidth]{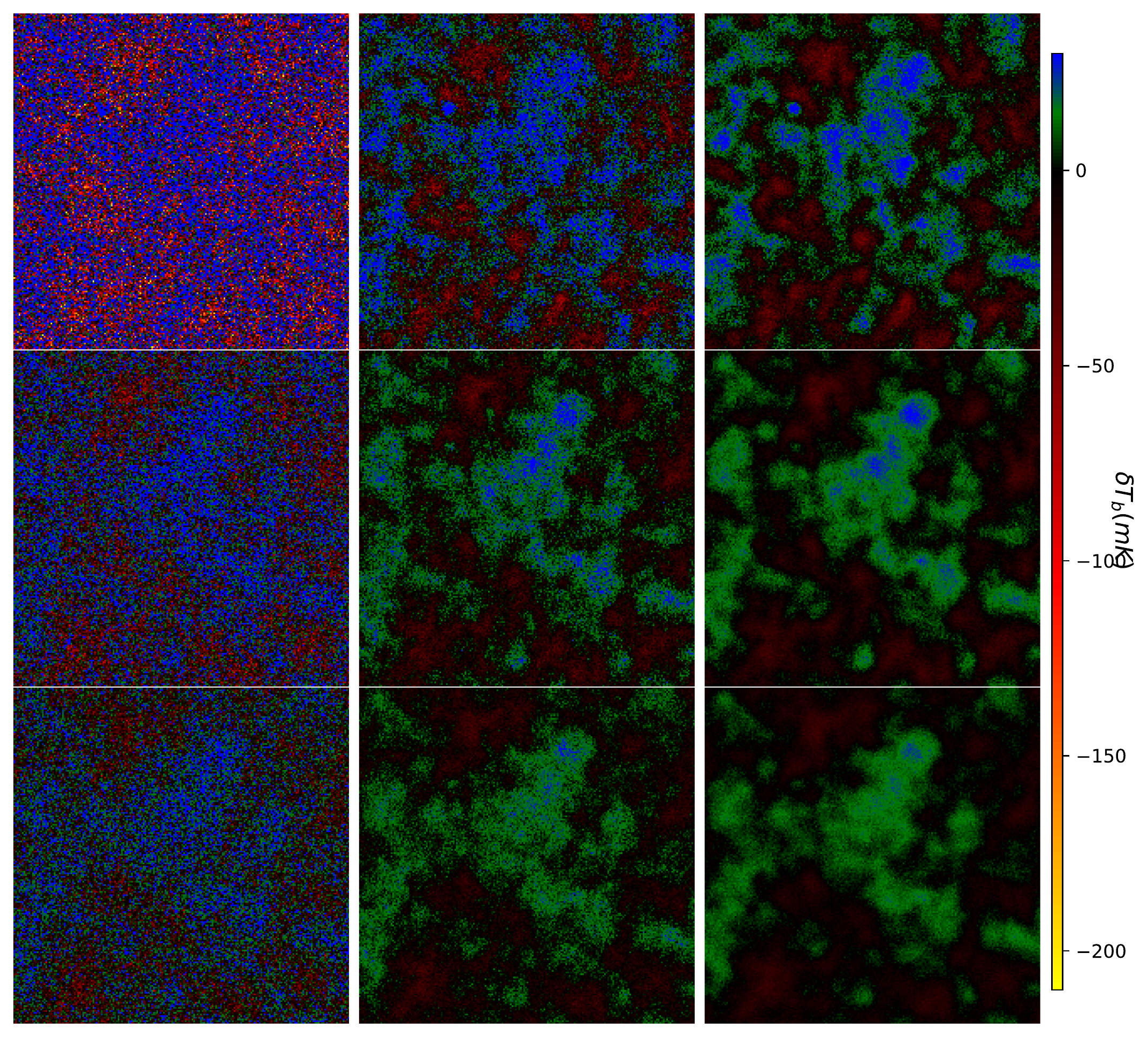}
\caption{Same as Figure~\ref{fig:2.3}, but for the ``noisy'' 21-cm maps after adding white Gaussian noise by assuming 1000-hour integration time of SKA.}
\label{fig:2.6}
\end{figure}

Figure~\ref{fig:2.6} further shows how adding noise affects the same mock 21-cm maps to Figure~\ref{fig:2.3}, by assuming 1000-hour SKA integration time. 
At large beamsize and frequency bandwidth (e.g., bottom-right panel of Figure~\ref{fig:2.6}), the RMS value of the smoothed mock 21-cm maps is higher than the sensitivity limit.
Also, the overall distribution of the noisy 21-cm map is similar to the smoothed 21-cm map.
On the other hand, at small beamsize and frequency bandwidth (e.g., top-left panel of Figure~\ref{fig:2.6}), the RMS value is about an order of magnitude smaller than the sensitivity limit.
As a result, sophisticated analysis methods that rely the detailed spatial distribution of $\dTb^{\rm n}$ might find difficulties in such maps without an additional denoising process.

Note that, however, one can see a visual pattern of patches with generally positive and negative values of $\dTb^{\rm n}$ at the top-left panel of Figure~\ref{fig:2.6}, even before any denoising process.
Also, one can see that such visual pattern has similarity to the clearer patterns shown at other panels (i.e., noisy 21-cm maps with larger beamsize and frequency bandwidth).
However, although there exist some candidates such as binning or Gaussian smoothing, the mathematical definition of such ``visual pattern'' may not be straightforward.
Note that such recognition of mathematically unclear visual patterns is one of the strongest specialties of deep learning technique, which we will see in the next section.

\section{Deep Learning Prediction of Mean Neutral Hydrogen Fraction}\label{sec:3}
\subsection{Deep Learning Architecture}

\begin{figure}
\centering
\includegraphics[width=\textwidth]{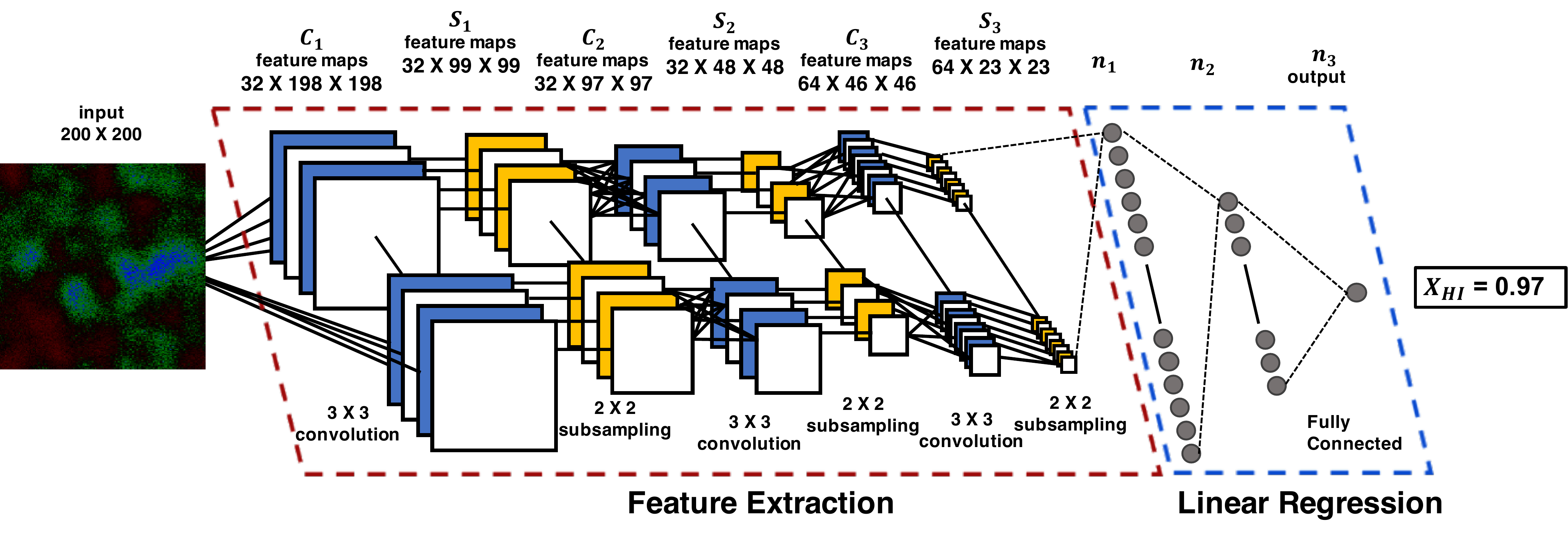} \\ \vspace{20pt}
\includegraphics[width=\textwidth]{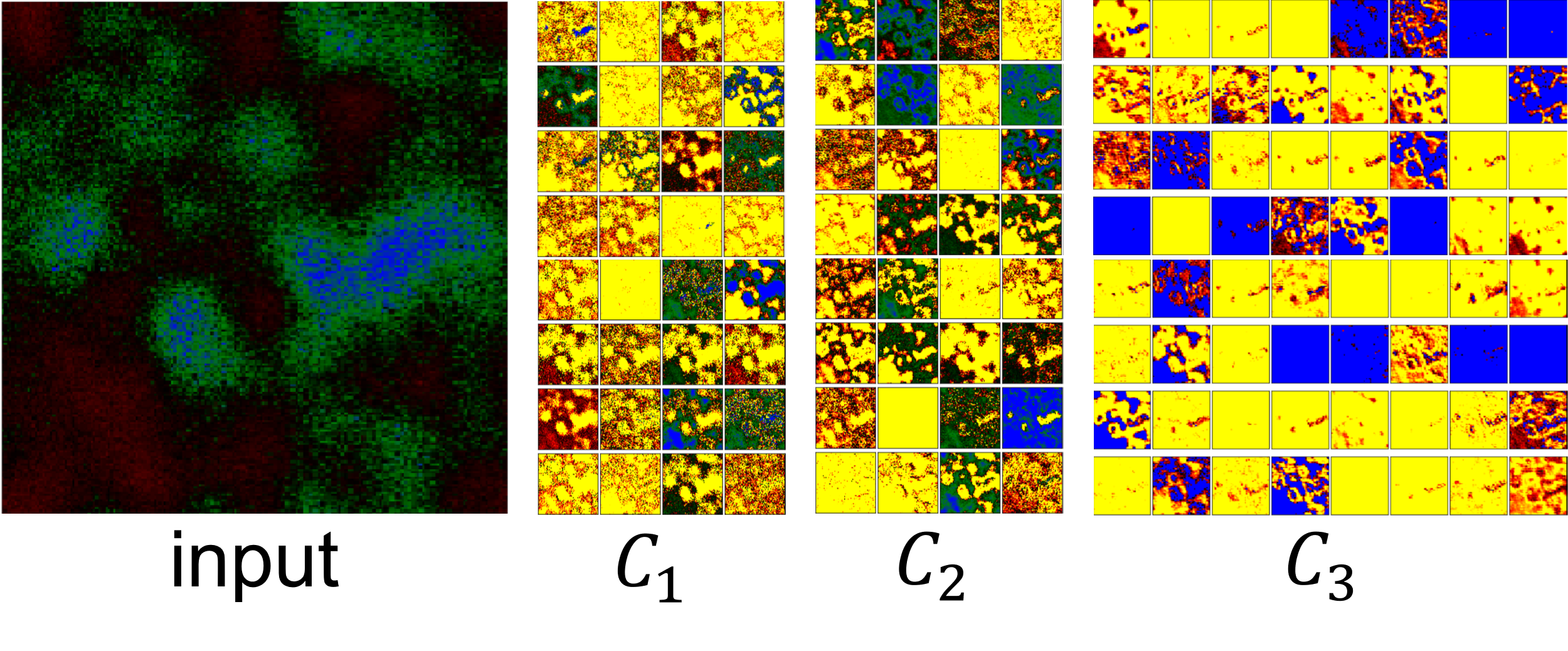}
\caption{Top panel: schematic architecture of convolutional neural network used for training. 
Bottom panel: layer-by-layer outputs of an example noisy 21-cm map input with $(z,\avgxHI,\Delta \theta,\Delta \nu, \Delta T) = (11.93, 0.97, 3', 0.2\MHz, 1000\,{\rm hours})$ during the feature extraction.}
\label{fig:3.2}
\end{figure}

\begin{table}
\caption{Outline of convolutional neural network used for training.}\label{tab:3.1}
\begin{ruledtabular}
\begin{tabular}{ccccc}
Layer & \# Filters & Filter Size, Stride & Padding & Output Dimension \\
\colrule
Input & - & - & - & (3, 200, 200) \\
Conv2D-1 & 32 & (3,3), (1,1) & valid & (32, 198, 198) \\
BatchNorm-1 (ReLU) & - & - & - & (32, 198, 198) \\
MaxPool2D-1 & 1 & (2,2), (2,2) & valid & (32, 99, 99) \\
Conv2D-2 & 32 & (3,3), (1,1) & valid & (32, 97, 97) \\
BatchNorm-2 (ReLU) & - & - & - & (32, 97, 97) \\
MaxPool2D-2 & 1 & (2,2), (2,2) & valid & (32, 48, 48) \\
Conv2D-3 & 64 & (3,3), (1,1) & valid & (64, 46, 46) \\
BatchNorm-3 (ReLU) & - & - & - & (64, 46, 46) \\
MaxPool2D-3 & 1 & (2,2), (2,2) & valid & (64, 23, 23) \\
\colrule
Flatten & - & - & - & 33856 \\
FC-1 (He-uniform) & - & - & - & 64 \\
BatchNorm-4 (ReLU) & - & - & - & 64 \\
FC-2 (He-uniform) & - & - & - & 32 \\
BatchNorm-5 (ReLU) & - & - & - & 32 \\
FC-3 (linear) & - & - & - & 1 \\
\end{tabular}
\end{ruledtabular}
\end{table}

We use the convolutional neural network (CNN) optimized to recognize visual patterns within the image dataset with efficiently lessening computational cost for expressing complex non-linear function for multi-dimensional input data.
Figure~\ref{fig:3.2} and Table~\ref{tab:3.1} show our CNN architecture design.
The entire network is divided into two parts, feature extraction and (fully-connected; FC) linear regression. 
All the filters in the first part of the architecture have the same size, $(3, 3)$, and are employed to extract the feature of the processed 21-cm map through a series of convolutions. Before each convolution layer, we use the max pooling (MaxPool) layer. The MaxPool layer helps to ease operations needed over the training by reducing the features resulting from convolution one, and can improve the result by keeping the network from getting over-fitted with a represented value (c.f. maximum or mean) of particular features over from the local regions, without numerous feature maps more than necessary. Finally, after the convolution and pooling layers, it is followed by a linear regression part, which consists of three FC layers. The FC layers employ the high-level features from the first part of the CNN and help to learn the network be close to have a desired output value through the last fully-connected layer activated by a linear function. 

We apply the rectified linear unit (ReLU), which is defined as ${\rm ReLU}(s) = \max (0,s)$, as the activation functions of all convolutional and FC layers\cite{hahnloser2000, glorot2011}.
The weights on the fully-connected layers are initialized by applying He uniform initialization\cite{c52}, which extract $n_{\rm in}$ samples from weights initialized from a uniform distribution between $-\sqrt{6/n_{\rm in}}$ and $+\sqrt{6/n_{\rm in}}$.
The He uniform initialization for weights technique helps in reaching a global minimum of the cost function faster and more effectively.

Before each ReLU activation, we also add batch normalization (BatchNorm) layers\cite{c50}. 
For the situation where the inputs are fed with forwarding to the output layer, the distributions of inputs keep changing with each iterations because the weights of each layers are changed over the training of network. So the intermediate layers are hard to adapt its changing distributions of inputs. 
To prevent such problems, BatchNorm renormalize a mini-batch of inputs ($x_i$)  into $y_i$ as:
\begin{equation}
y_i = \gamma \frac{x_i - \mu}{\sqrt{\sigma^2 + \epsilon}} + \beta \, ,
\end{equation}
where $\mu$ and $\sigma$ are the mean and standard deviation of $x_i$, $\epsilon$ is a small constant value introduced to avoid divergence, and $\gamma$ and $\beta$ are two additional learnable parameters introduced in the BatchNorm process.
Not only BatchNorm can effectively complement some chronic problems that deep neural network has, such as gradient vanishing problem, but allow us to use much higher learning rate and perform as a regularizer for a part.


\subsection{Training}

For the CNN training, first of all, we produce the samples by splitting the noisy 21-cm maps with $200^3$ grids into 67 slices of $200 \times 200$ pixels along the LOS direction.
We calculate the mean neutral hydrogen fraction for each slice.
Note that, while we only use $18\sim 35$ snapshots between $z = 6 \sim 13$, $\avgxHI$ estimated from different slices at the same redshift may vary.
As a result, for a given $(\Delta \theta, \Delta \nu, \Delta T)$, we have $1206 \sim 2345$ samples for each boxsize and high-resolution grid setup. 
Note that the difference of angular field of view from a fixed boxsize between $z = 6 \sim 13$ is less than $13\%$, which we will consider as nearly identical.
We have also found that the performance is similar even if one uses 10318 samples from all 9 boxsize- and high-resolution-grid-setups, and therefore, we use all 10318 samples hereafter.

\begin{table}
\caption{Colormap scheme used to generate input images from the noisy mock 21-cm maps.
Colors for $\dTb^{\rm n}$ values in between are determined with the linear interpolation.}\label{tab:colormap}
\begin{ruledtabular}
\begin{tabular}{ccccc}
$\dTb^{\rm n}$ [$\mK$] & Color Name & Red [$0\sim 1$] & Green [$0\sim 1$] & Blue [$0\sim 1$] \\
\colrule
$-210$ & Yellow & 1 & 1 & 0 \\
$-105$ & Red & 1 & 0 & 0 \\
0 & Black & 0 & 0 & 0 \\
15 & Green & 0 & $0.502$ & 0 \\
30 & Blue & 0 & 0 & 1 \\
\end{tabular}
\end{ruledtabular}
\end{table}

We produce the inputs of our CNN architecture by converting each $200 \times 200$ pixels of $\dTb^{\rm n}$ slice into a 3-channel RGB image (i.e., an array with $(3,200,200)$ size) by applying a custom colormap scheme (see Table~\ref{tab:colormap} for definition and Figure~\ref{fig:extreme} for examples).
We have also tested using the $200 \times 200$ slice directly as input and found no notable difference in performance.

We then split our samples into three sets: training set used for training the model parameters, validation set used for checking the training process, and test set for evaluating the performance of the trained model.
For a given $(\Delta \theta, \Delta \nu, \Delta T)$, we split 10318 samples to 7428 training samples, 826 validation samples, and 2064 test samples.
We randomly split the samples into three sets while keeping the distribution of $\avgxHI$ similar.

We compile our CNN model with Adam optimizer\cite{c53} with a learning rate $10^{-3}$. 
Also, we adopt the mean squared error (MSE) as the loss function,
\begin{equation}
{\rm MSE} = \frac{1}{m} \sum_{i=1}^{m} \left( \avgxHIsub{pred}^i - \avgxHIsub{true}^i \right)^2 \, ,
\end{equation}
Here, $m$ is the minibatch size for training, which is set to 8.
We have also tested several minibatch sizes between 4 and 64 and found no notable difference in performance for $m \geq 8$.

We define an epoch as an iteration of fitting the model parameters from 7428 training samples and evaluating its training loss as well as validation loss from 826 validation samples.
Since we set the minibatch size to 8, the number of minibatches per epoch is 929.
For a single run of CNN training, we set the maximum number of epochs as 500. However, for saving the time and preventing the overfitting, we stop the training when the validation loss is not improved for 30 epochs, and the typical number of epochs becomes $\sim 200$.
We then use the model at the epoch where the validation loss is minimized.
We perform our training with Keras\cite{chollet2015} with Tensorflow GPU version as a backend\cite{abadi2015}, and each run takes $\sim 2$ hours with a single NVIDIA V100 GPU card in a NVIDIA DGX-1 GPU platform.

Furthermore, we make use of the cross validation (CV) technique to test the stability of a certain deep learning model. In CV technique, the entire dataset is split into $K$-folds, and multiple CNN models are trained by using a certain fold as validation set and the remaining $(K-1)$-folds as training set.
We use the shuffled 10-folded CV so that we train our network 5 times with the number of epoch and data having different distribution for each time.
We have found that the validation loss from different models agrees within less than $75\%$ deviation from their average.

\section{Results}\label{sec:4}

The bottom panel of Figure~\ref{fig:3.2} shows how our CNN model processes the input noisy 21-cm map during the feature extraction.
Numerous feature maps are similar to the black-and-white segmented images of the input map with different intensity thresholds and small-scale noises.
It means that our CNN model allows a similar strategy to the most straightforward estimation method of $\avgxHI$, i.e., measuring the area fraction of zero $\dTb^0 (\vec{\theta})$.
Allowing different intensity thresholds is mainly because the value of $\dTb^{\rm s} (\vec{\theta})$ at $\dTb^0 (\vec{\theta}) = 0$ is uncertain, due to the renormalization of the mean value.
Also, allowing different small-scale noises may be related to denoising the noisy 21-cm map in several ways.
Note that the variations of intensity thresholds and small-scale noises depend on $(\Delta \theta, \Delta \nu, \Delta T)$.

\begin{figure}
\includegraphics[width=\textwidth]{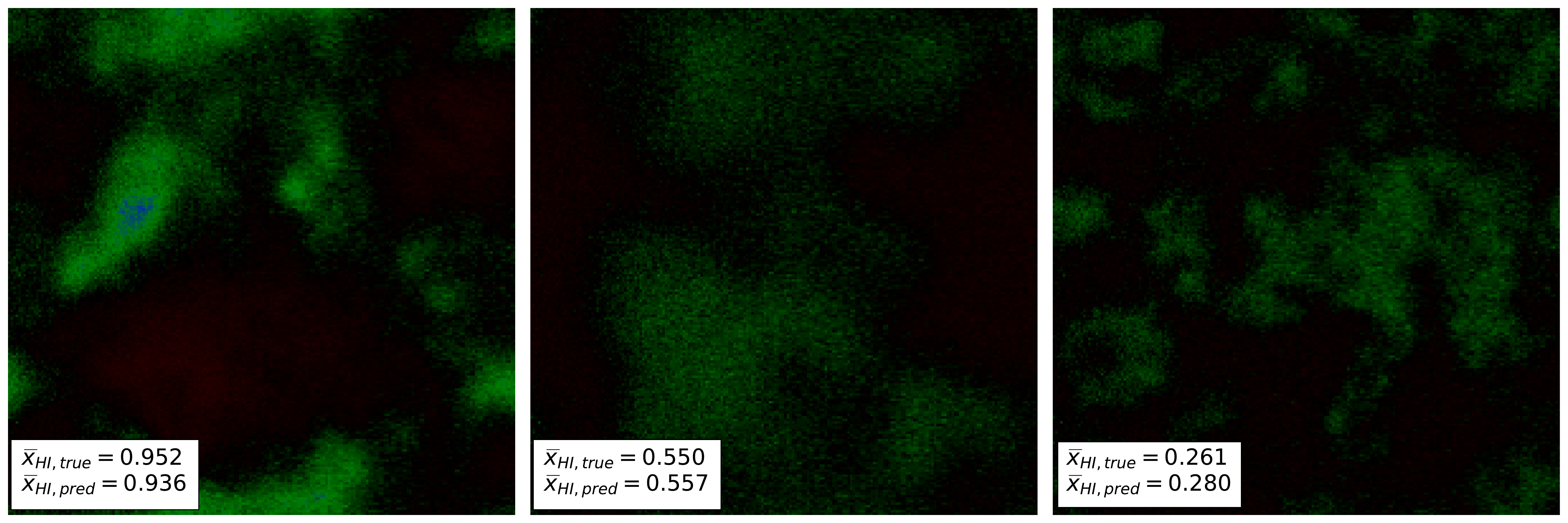} \\
\includegraphics[width=\textwidth]{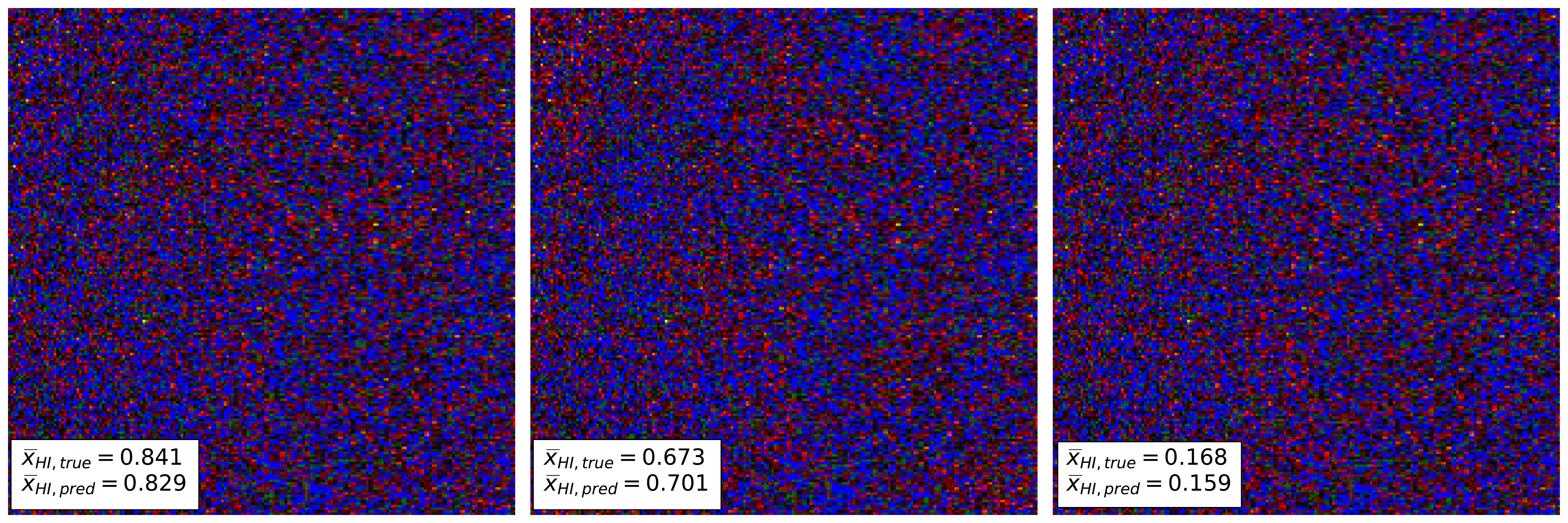}
\caption{The noisy 21-cm map, mean neutral hydrogen fraction, and its prediction from our CNN model for the two extreme choices of beamsize and frequency bandwidth.
Top panel: extremely smoothed signal with $(\Delta \theta, \Delta \nu) = (3', 2\MHz)$.
Bottom panel: extremely noisy signal with $(\Delta \theta, \Delta \nu) = (1', 0.2\MHz)$.
We assume 1000-hour integration time of SKA.}\label{fig:extreme}
\end{figure}

Figure~\ref{fig:extreme} shows the noisy mock 21-cm maps for the two extreme choices of beamsize and frequency bandwidth.
The top panel of Figure~\ref{fig:extreme} corresponds to large values of beamsize and frequency bandwidth.
Although the observational noise is relatively small, the entire 21-cm map is too smoothed so that the 21-cm maps from three characteristic epochs ($\avgxHI \sim 1$, $0.5$, and $0$) looks similar.
On the other hand, the bottom panel that comes from small $(\Delta \theta, \Delta \nu)$ has more observational noise than the signal itself, which also makes the 21-cm maps from three epochs hard to visually distinguish.
Interestingly, our CNN method can clearly distinguish such noisy 21-cm maps at different epochs with the prediction error of $\avgxHI$ less than $\sim 0.03$.

\begin{figure}
\includegraphics[width=\textwidth]{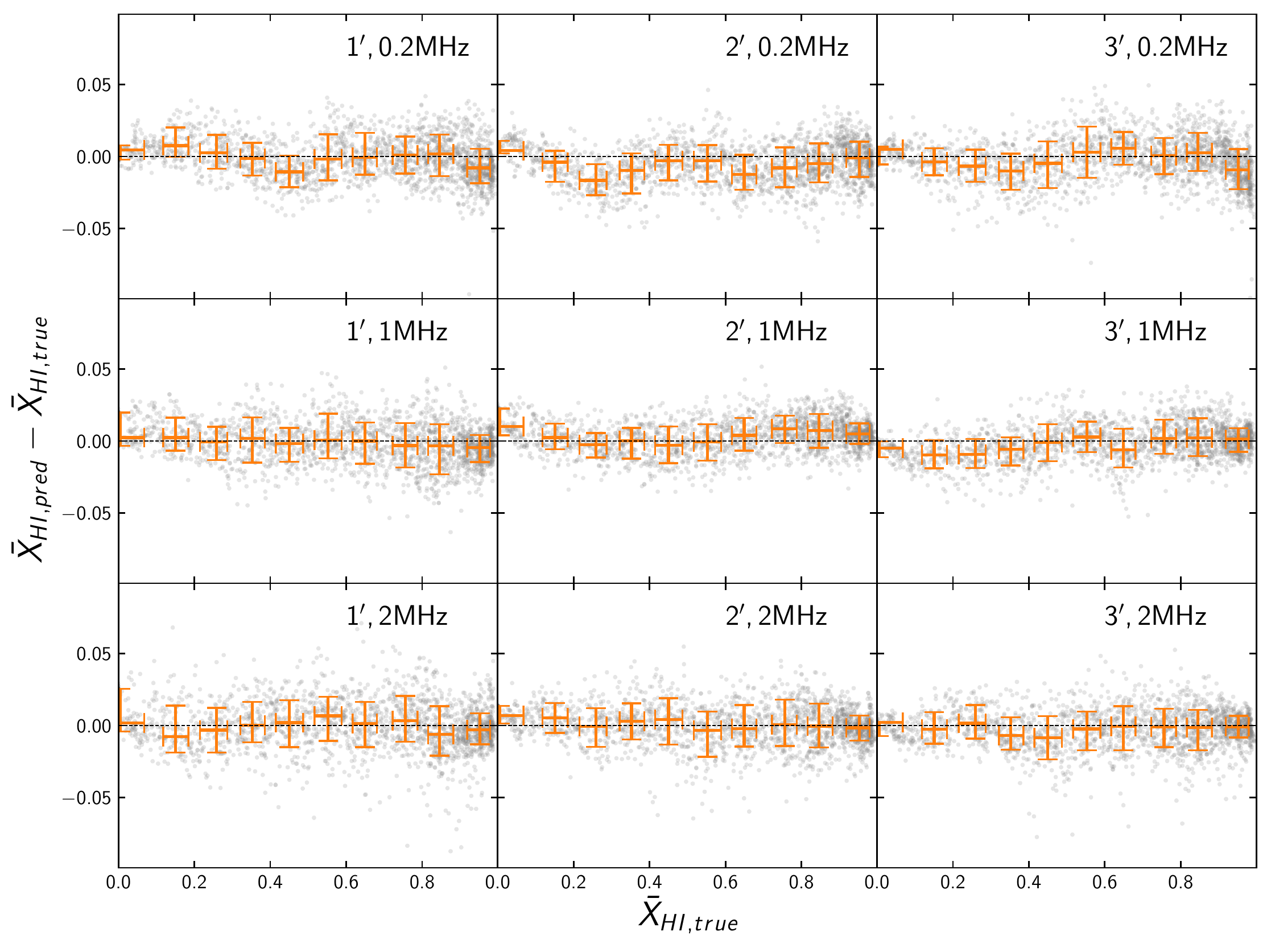}
\caption{Prediction error of the mean neutral hydrogen fraction from our CNN method as a function of its truth value.
Dots: 2064 test samples.
Error bars: median and 68\% certainty level from 10 equally spaced bins of $\avgxHIsub{true}$.
From left to right: $\Delta \theta = 1'$, $2'$, and $3'$.
From top to bottom: $\Delta \nu = 0.2\MHz$, $1\MHz$, and $2\MHz$.
We assume 1000-hour integration time of SKA.}
\label{fig:1000h}
\end{figure}

\begin{table}
\caption{Summary of the absolute prediction error $|\avgxHIsub{pred} - \avgxHIsub{true}|$ from our CNN method.
We assume 1000-hour integration time of SKA.
The best and worst cases are marked as \textbf{bolded} and \underline{underlined}, respectively.}
\label{tab:1000h}
\begin{ruledtabular}
\begin{tabular}{cc|cccc}
Beamsize & Bandwidth & Median & $1\sigma$-high & $2\sigma$-high & Maximum \\
\colrule
\multirow{3}{*}{$1'$} & $0.2\MHz$ & 0.0090 & 0.0198 & 0.0409 & 0.0956 \\
& $1\MHz$ & 0.0087 & 0.0197 & 0.0474 & 0.0633 \\
& $2\MHz$ & \underline{0.0100} & \underline{0.0227} & \underline{0.0687} & 0.0871 \\
\colrule
\multirow{3}{*}{$2'$} & $0.2\MHz$ & 0.0092 & 0.0202 & 0.0490 & 0.0590 \\
& $1\MHz$ & 0.0082 & 0.0169 & \textbf{0.0349} & \textbf{0.0517} \\
& $2\MHz$ & 0.0082 & 0.0184 & 0.0544 & 0.0643 \\
\colrule
\multirow{3}{*}{$3'$} & $0.2\MHz$ & 0.0091 & 0.0212 & 0.0504 & \underline{0.0980} \\
& $1\MHz$ & 0.0075 & \textbf{0.0162} & 0.0429 & 0.0524 \\
& $2\MHz$ & \textbf{0.0075} & 0.0178 & 0.0616 & 0.0772 \\
\end{tabular}
\end{ruledtabular}
\end{table}

Figure~\ref{fig:1000h} and Table~\ref{tab:1000h} show the prediction power of our CNN method for different choices of beamsize and frequency bandwidth by assuming 1000-hour integration time of SKA.
Overall, the predicted mean neutral hydrogen fraction has a good agreement of its truth value.
In the case $(\Delta \theta, \Delta \nu) = (2', 1\MHz)$, the median value of the absolute prediction error $|\avgxHIsub{pred} - \avgxHIsub{true}|$ is about $0.008$, while the upper bound of its 95\% certainty level is about $0.05$.
Note that such accuracy on reconstructing history of $\avgxHI(z)$ is comparable to the forecast presented in \cite{liu2016,park2019}.
Therefore, although it is beyond our scope, one could expect that our CNN method might provide a similar constraint power to the reionization parameters.

While our CNN method can predict $\avgxHI(z)$ reasonably well with various choices of beamsize and frequency bandwidth in overall, there exists non-negligible difference of performance between them.
For example, configurations with small beamsize and frequency, especially those with $\Delta \nu = 0.2\MHz$, tend to have a systematic bias with a wavy shape.
This might mean that, due to the significant noise level, the CNN model tends to behave a bit similar to classification rather than pure regression.
On the other hand, those with large beamsize and frequency, especially with $\Delta \nu = 2\MHz$, tend to have a large scatter from zero at the tail of probability distribution.
This could be possible if such outlier maps contain many small ionized bubbles (or the small tip of bubbles) that contribute non-negligible fraction to $\avgxHI$.

\begin{figure}
\includegraphics[width=0.8\textwidth]{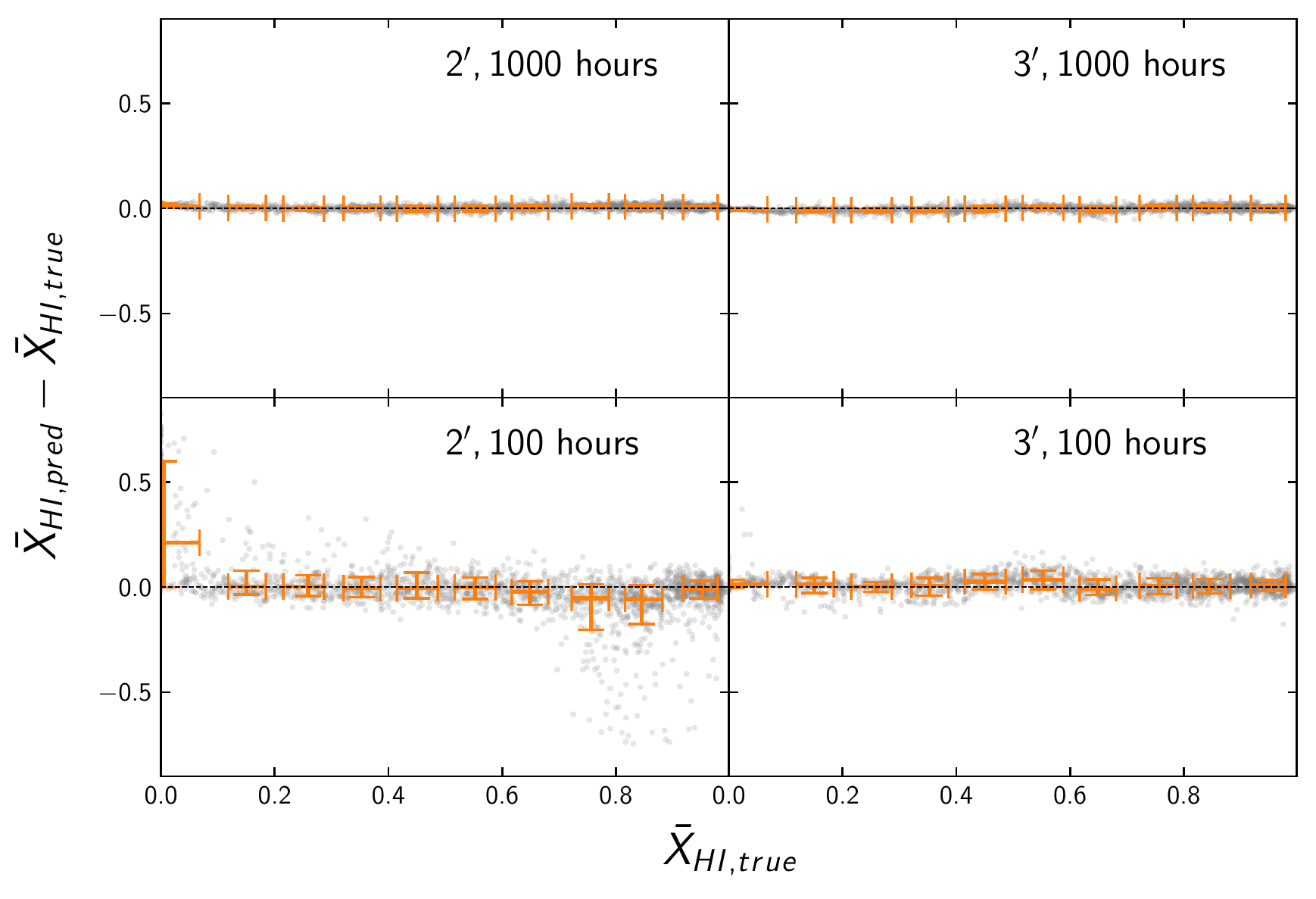}
\caption{Same as Figure~\ref{fig:1000h}, but by varying $(\Delta \theta, \Delta T)$ while fixing $\Delta \nu = 1\MHz$.
From left to right: $\Delta \theta = 2'$ and $3'$.
From top to bottom: $\Delta T = 1000$ hours and 100 hours.}
\label{fig:1mhz}
\end{figure}

\begin{table}
\caption{Same as Table~\ref{tab:1000h}, but by varying $(\Delta \theta, \Delta T)$ while fixing $\Delta \nu = 1\MHz$.}
\label{tab:1mhz}
\begin{ruledtabular}
\begin{tabular}{cc|cccc}
Beamsize & Integration Time & Median & $1\sigma$-high & $2\sigma$-high & Maximum \\
\colrule
\multirow{2}{*}{$2'$} & 1000 hours & 0.0082 & 0.0169 & \textbf{0.0349} & \textbf{0.0517} \\
& 100 hours & \underline{0.0416} & \underline{0.118} & \underline{0.746} & \underline{0.821} \\
\colrule
\multirow{2}{*}{$3'$} & 1000 hours & \textbf{0.0075} & \textbf{0.0162} & 0.0429 & 0.0524 \\
& 100 hours & 0.0227 & 0.0530 & 0.163 & 0.367 \\
\end{tabular}
\end{ruledtabular}
\end{table}

In Figure~\ref{fig:1mhz} and Table~\ref{tab:1mhz}, we further test the prediction power of our CNN method with 100-hour integration time of SKA.
As the noise level is about an order of magnitude higher than the actual signal, there exists a significant amount of deviation even in large beamsize $\Delta \theta \gtrsim 3'$.
Furthermore, for $\Delta \theta \lesssim 2'$ it clearly shows two populations with $(\avgxHIsub{true}, \avgxHIsub{pred}) \approx (0, 1)$ and vice versa.
It means that the CNN model fails to distinguish the 21-cm maps before  and the late stage of EoR --- in other words, the number of ``classification'' category that our CNN model with large noise level tends to find becomes extremely low.
Note that the SNR at $(\Delta \theta, \Delta \nu, \Delta T) = (2', 1\MHz, 100\,{\rm hours})$ is similar to that at $(1', 1\MHz, 1000\,{\rm hours})$, where the prediction accuracy of $\avgxHI$ is reasonably good.
It emphasizes that, since the distribution of 21-cm map is highly non-Gaussian in general, the SNR (or, $\dTbrms/\sigma$) alone cannot fully determine the prediction power of our CNN method.
Further study on the deep learning architecture or the combination with other analysis methods might be helpful to enhance the prediction power of $\avgxHI$, which is beyond our scope.

\section{Conclusions}\label{sec:5}

In this paper, we introduced a novel convolutional neural network (CNN)-based deep learning technique for the prediction of the mean neutral hydrogen fraction ($\avgxHI$) during the Epoch of Reionization (EoR) from the two-dimensional tomography of the redshifted 21-cm maps of differential brightness temperature. 
We first simulated the undistorted 21-cm maps of $200^3$ uniform grids with $50\sim 300\,{\rm cMpc}$ boxsize between $z = 6 \sim 13$ by using a semi-numerical simulation code 21cmFAST. 
We then applied various instrumental conditions to the dataset by controlling beamsize and frequency bandwidth suitable for the upcoming Square Kilometre Array (SKA) to produce the noisy mock 21-cm maps. 
After converting the noisy mock 21-cm maps into RGB images, we applied them as inputs of our CNN architecture to predict the corresponding values of $\avgxHI$. 

The main results of this paper can be summarized as follows.
\begin{enumerate}
\item Our CNN method has a capability to predict $\avgxHI$ from the raw noisy 21-cm maps even when the overall signal-to-noise ratio (SNR) is less than unity, depending on the radio survey configuration.
It is because the 21-cm maps during the EoR is highly non-Gaussian in general, so that there exists many additional features than just SNR that the CNN method can utilize.

In a similar reason, the overall performance of our CNN method depends on the combination of radio survey configuration, such as beamsize, frequency bandwidth, and telescope integration time, rather than just SNR.

\item For survey configuration with low-SNR (e.g., ${\rm SNR} \lesssim 1$), there exists a systematic bias with a wavy shape between the CNN prediction of $\avgxHI$ and its truth value.
This might mean that the CNN method becomes closer to the classifier of discrete categories rather than regressor of continuous value, mainly due to the high noise level.

On the other hand, survey configuration with large smoothing (e.g., large beamsize and frequency bandwidth) may allow large scatter of prediction error, partly due to the slices containing large number of small ionized bubbles.

\item If one uses 1000-hour integration of SKA, beamsize and frequency bandwidth $\Delta \theta \simeq 2'\sim 3'$ and $\Delta \nu \simeq 1\MHz$ can be regarded as an optimal configuration for the prediction of $\avgxHI$ with the CNN method.
With such configuration, median and the $2\sigma$-upper bound of the absolute prediction error $|\avgxHIsub{pred} - \avgxHIsub{true}|$ are $\sim 0.008$ and $0.04$, respectively.
\end{enumerate}

Although our work can successfully reconstruct the reionization history during the EoR by incorporating some of the difficulties from actual radio observations, we have found there exists plenty of rooms for improvement.
For example, for more realistic performance test, it might be useful to adopt a detailed information of beam shape and noise addition from future radio surveys, as well as Galactic and extragalactic HI foreground\cite{asorey2020}.
Also, including 21-cm maps with varying cosmological and reionization parameters would be helpful to understand how our reconstruction of $\avgxHI(z)$ depends on such parameters.

\begin{acknowledgments}
The authors thank Kyungjin Ahn, Hyunbae Park, Dongsu Bak, Sangnam Park, David Parkinson, Jacobo Asorey, and an anonymous reviewer for helpful discussion and comments.
The authors were supported by Basic Science Research Program through the National Research Foundation of Korea funded by the Ministry of Education (2018\-R1\-A6\-A1\-A06\-024\-977).
Computational data were transferred through a high-speed network provided by the Korea Research Environment Open NETwork (KREONET).
\end{acknowledgments}

\bibliography{article}
\removed{

}

\end{document}